# An experimental and numerical study on the behavior of finite-length column vortex

Swetarka Das[1], Dharmendra Yadav[1], Abhisek Kumar Gupta[1], Debopam Das[1,2], Ashoke De[1,2,a]

[1]*Department of Aerospace Engineering, Indian Institute of Technology Kanpur, 208016, Kanpur, India*

[2]*Department of Sustainable Energy Engineering, Indian Institute of Technology Kanpur, 208016, Kanpur, India*

This paper explores experimental and numerical investigation of the spatio-temporal dynamics of a finite-length vortex column in three dimensions using Particle Image Velocimetry (PIV) and Large Eddy Simulation (LES). The research examines the combined impact of bending, buckling, and core-splitting on a finite-length vortex column. More precisely, the work centers on the fundamental motions, evolutions in flow along the axis, how the shape of the vortex core changes over time, the instability caused by the long waves on the vortex column, and the division of the vortex core. A novel prototype is developed that utilizes piston-driven inflow and produces a vortex column through the principles of flow separation and Biot-Savart induction, which is studied for predicting an ex-situ cyclonic line vortex. The present study discusses the complete three-dimensional overview of the same columnar vortex. The meridional swirl and the axial transport of vorticity deform the shape of the core cross-sections in different lengths of the column. The jump in the axial velocity at the core boundary allows for the occurrence of bending instabilities with a left-handed helical structure. These instabilities have a long wavelength, similar to the length of the vortex column, and belong to the m=+1 mode. The vortex column experiences a non-uniform curvature and torsion caused by the intricate shape of the laboratory model and varying flow speeds at different heights of the vortex column. The results offer valuable insights into the role of inertia, Coriolis, and viscous forces on the dynamics of the vortex column.

___________________________

[a)] Author to whom correspondence should be addressed:  ashoke@iitk.ac.in

## I. INTRODUCTION

Vortex sheet and core duo have been a significant topic for scientific discussions for a long age. This particular physics plays a significant role in everyday life. A vortex sheet is a shear layer with mathematically discontinuous tangential velocity. This leaves a high vorticity at vortex sheet regions. In the practical world, the viscous diffusion effect of fluid smooths out such sharp discontinuities in the flow fields. Such vortex sheets are often noticed to curl with time to form the sheet and the core pair. The curling of the vortex sheet occurs due to the induction of velocity on the sheet by the effect of the sheet itself. This self-induced velocity is mathematically defined by the Biot-Savart formula and Birkhoff-Rott equation. However, this equation is accurate analytically for only inviscid flows. Birkhoff-Rott equation calculates the self-induced velocity for inviscid Eulerian problems. The viscosity brings the local rotationality of fluid elements into the picture, which is more challenging to deal with analytically. Seeking the numerical solution to the rolling up of the vortex sheet has been a great challenge and a significant topic for scientific exploration for a long time in the late 20's. Kaden[1] provided a similarity solution for the spiral structure of the inner part of a semi-infinite vortex sheet. Discretizing the vortex sheet for numerical simulation of future states of the vortex sheet rolling up phenomenon became a challenge for the community. Rosenhead[2] and Westwater[3] implemented the point vortex discretization method for studying the evolution of the vortex sheet. Moore[4], Saffman and Baker[5] discovered the chaotic motion of the point vortices in this discretizing algorithm at the inner end of the vortex sheet when spatial and temporal resolutions were little refined. Moore introduced a tip vortex representing the inner part of the vortex sheet as a combined unit, which is collected from Smith's[6] idea. As the discretized point vortices approach the spiral's center, they dump their strength into the defined tip vortex. This phenomenon is referred to as 'Core Dumping'. Fink and Soh[7], and Baker[8] corrected this method while considering the mathematical and geometrical errors contributing to dynamic sheet roll-up's erroneous results. Several other methods (Hoeijmakers and Vaatstra[9]; Sugioka and Widnall[10]; Baker and Pham[11]; Higdon and Pozrikidis[12]; Sohn, Yoon, and Hwang[13]) were developed in different corners of the world that led to the proper analytic solutions of the vortex sheet roll-up phenomena. DeVoria et al.[14] discussed all this literature in detail in their recent publication.

The presence of viscosity brings shear and body rotation to the elemental level. Local vorticity is generated due to such local diffusion. Locally defined sharp shear layers, such as vortex sheets, get locally diffused in the normal direction of the sheet. Viscous diffusion brings difficulty in analytical formulation for developing the vortex sheet and the core at the center of the spiral. The tip vortex or the vortex core, in the case of the viscous flow condition, has been a primary concern for the scientific community for a long time. Rankine vortex core combines solid body rotation and a potential flow region outside the core boundary. Such a columnar vortex with solid body rotation type flow within and linear profile of azimuthal flow velocity is stable to any small perturbations. In 1880, Kelvin[15] showed the Rankine vortex was also stable in such small perturbations. Chandrasekhar[16] and Krishnamoorthy[17] showed that instability may develop if an axial flow within the columnar vortex exists. Temporal instability in vortices was investigated by many people (Moore and Saffman[18]; Uberoi *et al.*[19]; Lessen *et al.*[20]; Drazin and Reid[21]; Saffman[22]). Loiseleux *et al.*[23] studied the spatio-temporal instability in the presence of vortex breakdown. Loiseleux *et al.* investigated the effect of superimposing plug flow axial velocity profile with Rankine vortex on the growth of perturbations. The swirl and axial flow combination shows the behavior of vortex breakdown above a certain critical swirl level (Billant *et al.*[24]). These inviscid analytical analyses were based on linear stability analysis of spatio-temporal instability of Rankine vortex.

Instability waves grow in time and space, starting from a particular set of initial space and time perturbations. The source of perturbations can be the flow itself or the geometry. The interest never lies in how the perturbations are generated. Instead, which perturbations grow in time and space and how concludes the problem. These perturbations are the origin of turbulence in the flow field. Several authors discussed the interaction of these perturbations with the vortex core, i.e. the interaction of coherent structures (CS), the large scale vortices interact with the fine-scale turbulence. Zemn[25] found monotonic decay of turbulence around the vortex with time. The vortex keeps decaying at a constant rate. Wallin and Girimaji[26] demonstrated that the growth of decay of initial turbulence depends on the initial turbulence intensity. High-intensity turbulence decays and low-intensity turbulence grows in time. Turbulence approaches an equilibrium state that is not dependent on the initial condition. Experimental observations of Bailey and Tavoularis[27] indicated that a vortex responds to a turbulent excitation of a specific set of waves. Conversely, Pradeep and Hussain[28] explained that direct numerical simulation (DNS) results in the non-monotonic viscous decay of the turbulent vortex and the absence of any equilibrium mentioned in the above state. Perturbations that maximize the energy amplitude attain up to three-order-of-magnitude amplification at moderate Re (Antkowiak and Brncher[29,30]). This explains why bending waves are present in turbulent vortex cores (Melander and Hussain[31]). The evolution of such perturbations leads to a transition of the core and sustained turbulence surrounding the vortex core.

Another critical feature affecting a vortex column's dynamics is column buckling. The name "buckling" comes from the analogy to the buckling of solid columns under compressive axial load. The buckling phenomenon is quite famous in fluid mechanics problems (Taylor[32], Buckmaster[33], Cruickshank and Munson[34]). A Numerical investigation of Rogers and Moin[35] on turbulent flow under plain strain conditions sets one such example of buckling phenomena. In the case of in-plane strain conditions, flow is stretched in one direction and compressed in another. Hairpin vortices undergo buckling when they experience compression in the flow field. Such buckling phenomena lead to the breakdown of the vortices. Buckling phenomena are also reported in small-scale turbulent structures (Lundgren[36], Everson, and Sreenivasan[37]). Surrounding turbulent flow provides an environment of random compressions in the flow field. If the compression prevails for an extended length of the vortex axis, the column undergoes buckling. Marshall[38], developed a theory for critical conditions for helical instability of vortex columns due to axial compressive flow load. A vortex column isolated in an unbounded region is marginally stable to helical perturbations. If axial flow is introduced within the core only, it may lead to jet-like instability of the core (Widnall and Bliss[18,39], Moore and Saffman[18]). Marshall developed a linear theory on small amplitude helical waves on a columnar vortex in the presence of compressive flow load. His investigation led to a straightforward conclusion: if a vortex column has its '$k\sigma$' (k is wave number and $\sigma$ is the core radius) more than a critical value (approximately 1.77), the vortex will remain columnar, and if '$k\sigma$' is less than the critical value, the vortex will buckle. Linear stability analysis predicts finite maximum bending amplitude for a vortex column. The non-linear effects mainly govern large amplitude of helical disturbances and vortex breakdown.

Established literature describes the inviscid linear stability theory on vortex columns, buckling theory, linear helical instability theory, and vortex merging or splitting theory. These theories and models on infinite columnar vortex are following to the ideal. The vortex end effects do not interfere with the results, and non-linear effects are assumed to be absent. These theories and models accurately study the theoretical behavior of an ideal columnar vortex or a vortex filament. However, in a practical scenario, vortices encounter several such complicated phenomena altogether, along with spontaneous excitations due to surrounding turbulence. The practical dynamic behavior of a vortex depends on the superimposition of such basic dynamics (buckling, bending, etc.). The studies on columnar vortices with infinite length studies lack such practicality, and several essential dynamics, such as viscous dissipation effects, swirl-boundary layer interaction effects at the solid boundaries where the column vortex ends, etc. The finite length of a columnar vortex invites such complicated dynamics into the picture.

Given the preceding discourse, we believe it is imperative to examine the intricate dynamics of a finite-length vortex column subjected to a combined stimulation from the vortex's surroundings. Without a piston vortex or stopping vortex, the experimental setup covered in Das *et al.*[40] produces ring vortices. For this investigation, we have remained with the same experimental configuration. The setup makes advantage of an innovative design for the experimental model that creates a column vortex. An ex-situ illustration of a naturally occurring cyclonic vortex that moves downstream and intensifies as a result is provided by this study. To learn more about the path, strength, and temporal evolution of the vortex core as it moves downstream, one must examine the experimental results. A detailed study is conducted on the growth of this vortex on a cross-sectional plane at the mid-length of the vortex column. Predicting a cyclonic vortex's trajectory and strength progression, particularly that of a column vortex, requires specific information about the beginning circumstances, gradients of the flow (such as the pressure gradient and Coriolis forces), and other factors. The ensuing sections provide a detailed description of the vortex's genesis mechanism. A preliminary estimate of the instant positions and the vortex intensity at some discrete moments is extracted from the PIV data and experimental readings. Investigating the detailed physics underlying such a small-scale model of the cyclonic column vortex requires a three-dimensional overview of the column vortex. A comprehensive understanding of the dynamic behavior of a column vortex in perturbed surroundings can be gained by investigating all aspects of the three-dimensional dynamics of the column simultaneously through numerical research. The selected cases that comprise the superposition of all the three-dimensional phenomena stated above in the column vortex are presented in this study. The current study addresses the buckling of the vortex column as a result of variations in flow velocity along the column's length. Different dynamics of vortex evolution result from variations in the axial flow within the vortex column. Axial flow velocity distribution anisotropy causes complex column buckling. Helical disturbances grow in the vortex column with time and space because of the complex model design. For a specific time span, this paper offers a perfect case of the superimposition of vortex column buckling and bending instability. Moreover, the present study explores a critical feature of the vortex dynamics: the vortex core splitting. In the sections that follow, the physics underlying the splitting phenomena is covered in detail. The daughter cores that have split apart begin to loop around one another in a braided fashion. The third main topic of this article is the braided pattern and how it advances throughout time and space. This work examines five distinct scenarios based on five distinct Reynolds numbers (or inlet flow profiles). Completing the two-dimensional experimental observations requires a comprehensive three-dimensional overview of what happens to a column vortex when it is subjected to such complex dynamics. A precise method of predicting the ex-situ cyclonic line vortex in the current Laboratory model is demonstrated by the detailed results, their likely causes, and the superimposition of all complex dynamics that have been discussed above.

## II. PROBLEM STATEMENT

This article's primary goal is to examine the combined effects of vortex column buckling, bending, and splitting in a fully resolved three-dimensional scenario. These phenomena originate and are primarily affected by external stimuli and perturbations. Stimulus may originate due to the flow complexities or environmental influences. Small-scale flow disturbances occur primarily because of the complexity of geometries and case sets used in any experimental study. We discuss how a complex geometrical setup influences the flow and induces perturbations in the coherent structures. Such small-scale flow perturbations influence the dynamic evolution of the flow with time. However, the origin of perturbations in the vortex core due to complex geometries and complexities of flow dynamics do not fall in the scope of this work. Instead, this investigation focuses on the governing dynamics and observations regarding the dynamic evolution of the perturbed vortex core. In the present literature, a vortex is generated following the flow separation mechanism from a diverging solid surface. A sharp diverging turn of the solid boundary surface separates the flow stream from the solid boundary. If the flow is inviscid, the separation is clear, and a vivid shear layer develops between two sets of flow streams with different velocity directions. If the viscosity effect is notable (cases with less Reynolds number), the shear deformation and the viscous diffusion of fluid motion prevent the flow separation. In such a practical scenario, separation occurs if only the local inertia force is much larger than the local viscous forces. A flow separation from a triangular edge [41] (starting flow past an infinite edge[42]) originates a free shear layer that curls up due to the Biot-Savart induction of self-induced velocity of the vortex sheet and dumps the sheet strength into the vortex core at the center ("Core Dumping" [43]). This vortex core-sheet pair dynamically evolves and travels downstream with time.

Figure 1(a-b) presents the model "Cyclone Generator", which generates the column vortex that replicates a cyclone, hence the name. We have gathered experimental observations regarding the trajectory and evolution of the vortex column and performed numerical simulations of the same to validate our case. A thorough numerical investigation of these simulations leads us to a complete three-dimensional overview of the vortex column. Figure 2 portrays the schematic diagram of the model properly. The piston drives the air through a nozzle and dumps it into the main domain as a jet flow, as shown in the drawing representation (Figure 2). A sharp geometric turn (highlighted in Figure 2) introduces the flow separation, and a columnar vortex originates at that edge. This columnar vortex evolves and travels downstream with time. On the other side of the nozzle (opposite to the triangular edge side), the surfaces have several holes. Boundary layer roll-up and secondary

vortices develop on these surfaces during the flow time; these vortices clash with the study's objective column vortex and destroy it. These holes can introduce suction and blowing phenomena, delaying the flow separation and subsequent secondary vortex formation on these surfaces. The chimney is only intended to be used for this type of suction. When there is no interaction between the primary column vortex and the other nearby secondary vortices, the current study looks at the flow mechanics of the column vortex during a specific time range.

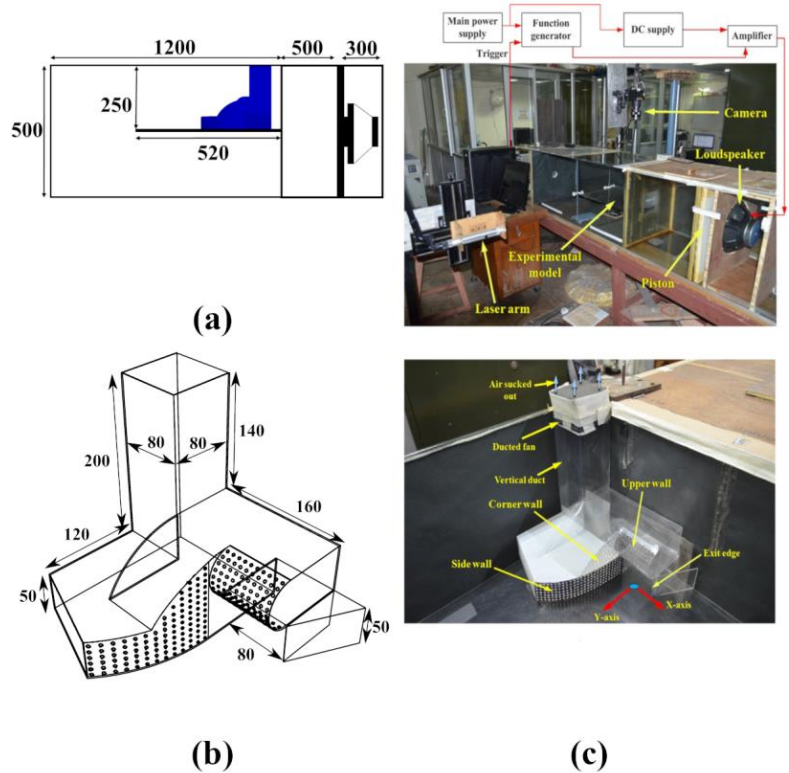

Figure 1: An overview of the (a) domain, (b) model, and (c) Experimental setup, (All dimensions are in mm)

Figure 1 presents the dimensions of the present geometry and the computational domain. The nozzle is 80 mm in width and 50 mm in height. The triangular edge, i.e. the sharp turn, encloses an angle of 26.56° with the jet direction (Figure 2). The triangular edge is blunt with a 0.5 mm tip width. The edge is as long as the height of the nozzle, i.e. 50 mm. The top surface at the nozzle exit gradually diverges, forming a semi-cylindrical structure with 10×11 (length×periphery) holes. The drilled holes are equidistant from each other and 3 mm in diameter and 1 mm in depth. The semi-cylinder has a radius of 30 mm and a length of 80 mm. The model is hollow inside, and the model walls are 1 mm thick. Holes are drilled on all surfaces where flow separation is inevitable except the surfaces of the triangular edge structure.

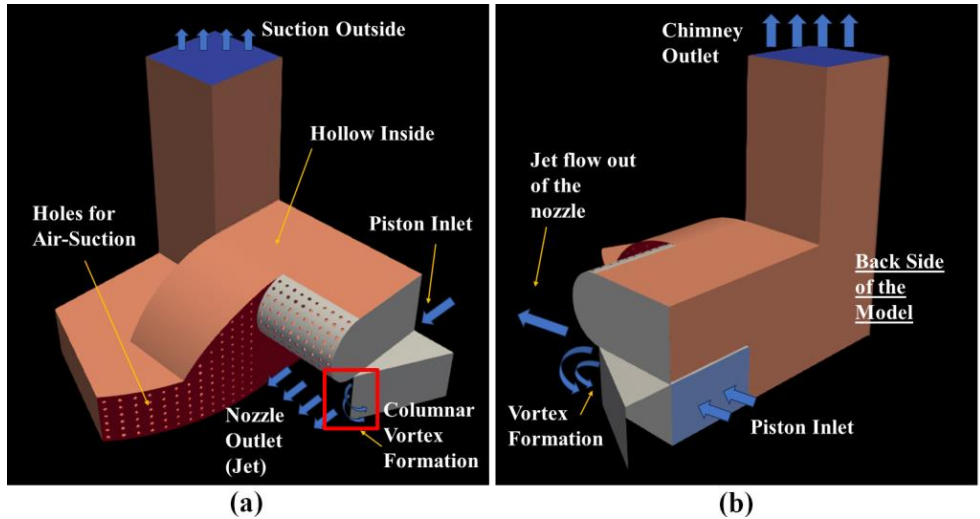

Figure 2: Schematic diagram for better understanding of the present model (front (a) and back (b) view)

Figure 4 shows the schematic representation of the entire computational domain and the boundary surfaces that enclose the fluid domain (refer to the boundary conditions). The computational domain does not include the piston-cylinder setup. The outlet of the piston-cylinder acts as the nozzle inlet in the present model. The piston drives the flow out of the cylinder, and the air enters the present model's nozzle. The rate of the air volume displaced by the piston is the same as the volumetric air flow rate at the piston cylinder exit (assuming incompressible fluid and flow). The incompressible flow continuity equation calculates the nozzle-inlet velocity using the area ratio of the piston-cylinder area to the nozzle area and piston velocity. The 'Initial and Boundary Conditions' subsection under 'Numerical details' throws some light on this topic. We have compared the numerical results against the measurements for completeness. Two principal planes are considered for this purpose, which cuts the vortex in the sagittal section and bisect the nozzle area. They are H1 and V1 planes, respectively, perpendicular to each other. Figure 3 shows these planes from an isometric view for the reader's better understanding.

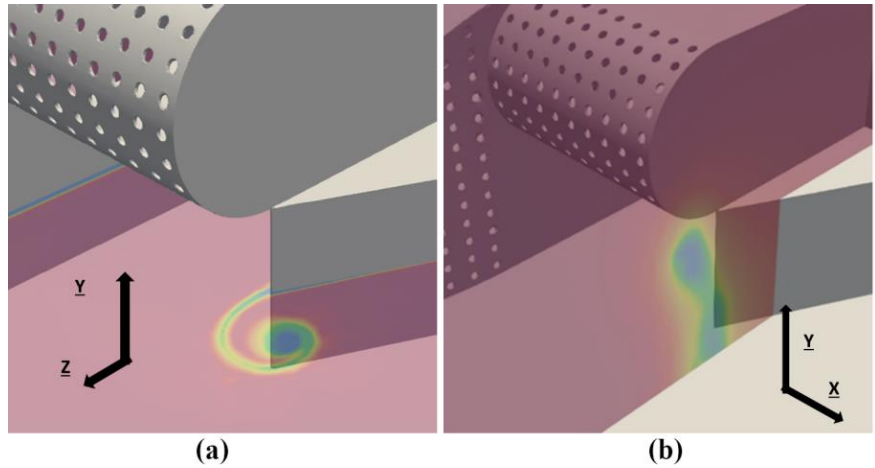

Figure 3: (a) H1 (y=25 mm) and (b) V1 (x=250 mm) plane in a three-dimensional isometric view of the model

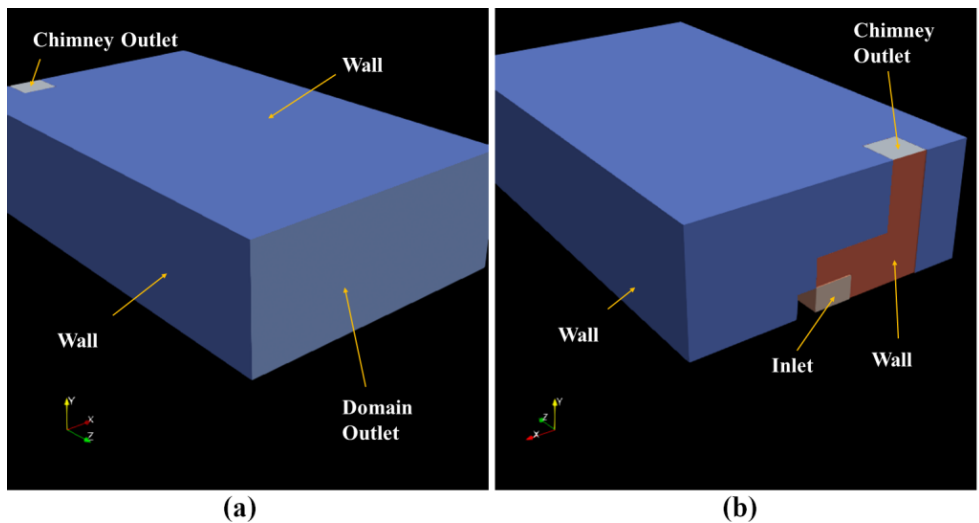

Figure 4: The surface boundaries of the computational domain

Since the geometry is complex, some important faces are tagged for the ease of further explanations in the coming discussions. Figure 5 presents the pictorial presentation of these surfaces. The triangular edge structure consists of two essential faces: face A and B. Face B is the blunt edge of the triangular wedge, and face A makes an angle of 26.56° with the nozzle flow direction (Z-direction of flow), as discussed earlier. Here, the width of face B is 0.5 mm. The nozzle outlet ends in a diverging surface, forming a semi-cylindrical radius of 30 mm (shown in figure 2). This face, namely face C, has several holes for suction and blowing phenomena, as discussed earlier. Face C promotes flow separation in the transverse plane (discussed in the following subsections). The boundary layer on this face rolls up as the flow velocity and the inertia force increase with time. This rolling up of the boundary layer creates secondary unnecessary vortices that disturb the objective vortex column. Holes and suction (blowing) phenomena help in delaying such scenarios. Face D is the top face of the nozzle. The face E is the straight surface starting from one of the sides of face C. Faces E and C play an essential role in vortex core splitting (described in later sections).

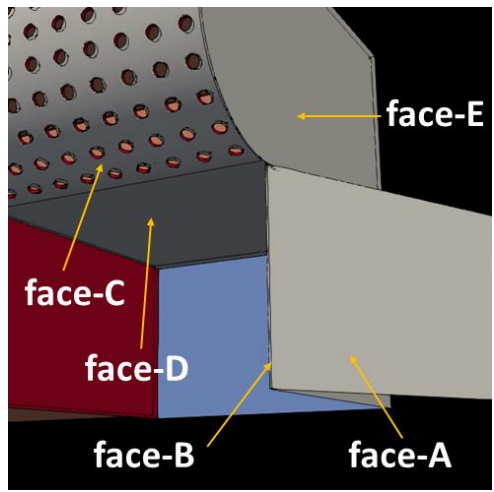

Figure 5: Important faces on the model wall

## III. EXPERIMENTAL SET-UP

A glass tank of 2 meters in length and 0.51 meters by 0.50 meters in cross-section makes up the basic experimental setup (see Figure 1-(a)). Two compartments are divided inside the glass tank: one holds the experimental model of the cyclone generator, and the other holds a loudspeaker piston. A plate with a thickness of 12 mm is used for compartmentalization, acquiring the same cross-sectional area as the tank. This wall, which separates two compartments, has an aperture of 0.135 m by 0.135 m. This aperture is used to introduce the experimental model. In the Cyclone Generator model, the only continuity between these two compartments is through the nozzle area. While the observations are made in the other compartment, we utilize the smaller compartment as the driver portion, which houses the piston. On the opposite end of the model is a piston located in the driver compartment. A flat, square-shaped, 25 mm-thick sheet of light polystyrene foam is used to construct the piston. The 100 W loudspeaker (SK12FRX) supplies the piston with impulsive movements. In addition to providing an adequate amount of ambient fluid, the larger cross-sectional area of the piston tank as opposed to the experimental model's nozzle also completely precludes the possibility of external air current during the experiment. Moreover, to reduce any unwanted vibrations during the piston motion, the experimental setup is firmly fastened atop a hardwood foundation. To ensure smooth piston movement without rubbing against the glass chamber wall, the piston plate area (0.50 m × 0.49 m) is slightly less than the glass chamber area. To guarantee a leak-proof setup, the little space between the piston and the wall is covered with polythene sheet strips. This study relies on the underlying assumption that the back pressure has not caused any appreciable deformation to the polythene strips. The polythene can displace up to 0.4% of the entire volume flow through the nozzle[40] when the piston moves.

The waveform generator (RIGOL DG-1022) provides signals to the loudspeaker that powers the piston, with a maximum amplitude of 20V. A Dual Tracking Power Supply PSD320 power amplifier is utilized to enhance the electrical signals produced by the function generator. This signal is amplified and sent to the loudspeaker. To achieve impulsive motion of the piston, which is necessary for vortex production, square waveforms with a period of 5 seconds are used to drive pistons (Didden[44], Maxworthy[45], Pullin[46]). At the top of the driving compartment, there is a square aperture measuring 0.20 m by 0.20 m that allows smoke to enter the compartment and serve as PIV seeding particles. An unfavorable pressure rise occurs in the experimental region as a result of mass slug transfer from the driving compartment to the experimental compartment. To prevent this kind of abrupt pressure surge, the vertical wall at the end of the experimental chamber features a 60 mm-diameter circular aperture. These H1 and V1 planes are where we visualize the flow fields that we have captured using Particle Image Velocimetry (PIV). Double-pulsed Quanta System laser (Nd: YAG laser, 200 mJ/pulse, 10Hz) is employed in this work. For laser 1, the ideal gap between the flash lamp's start and the Q-switch is 175 μs, while for laser 2, it is 225 μs. One millisecond separates the two pulses; this is chosen following trials at one, 1.5, and two milliseconds. The photographs are captured with an 8 MP, 12-bit CCD camera (equipped with 50 mm and 100 mm lenses). An eight-channel timer box (MicroPulse 725) synchronizes the camera and laser, and the waveform generator (RIGOL DG-1022) provides the external trigger. Since the available laser can send out pulses at a maximum frequency of 10 Hz, we use a phase-locking technique (with the same timer box) to acquire high time resolution. Every experiment is done with the same piston voltages (speed) and a 20 ms delay added to the laser and camera. Once we generate the images, we use PIVlab to perform PIV analysis. The PIV analysis is done in two steps: first with an interrogation window size of 128 × 128 pixels, then with a final pass of 64 × 64 pixels, with a 50% overlap between the windows. We illuminate the flow field for flow visualization studies using a continuous laser (1.2 W, 532 nm). We capture images with a high-speed camera that has a resolution of 1028 × 1296 pixels and a frame rate of 2 kHz at full resolution. The laser sheet is positioned in the horizontal plane H1 and the vertical plane V1 using a cylindrical lens, and images are captured at 200 frames per second (f.p.s).

## IV. NUMERICAL DETAILS

This section describes the computational components of this study. This section is further divided into five subsections. The 'Domain Discretization and Adaptive Mesh Refinement (AMR)' subsection focuses on the discretization of the computational domain for numerical simulations. The second subsection describes the initial and boundary conditions. The subsection under "Flow solver" covers the numerical solver and different numerical methods utilized in this work, while the fourth subsection discusses turbulent scale resolution and LES quality index. The final subsection presents the data validation for these numerical results with the experimental observations.

### A. Domain Discretization and Adaptive Mesh Refinement (AMR)

Physical domain (material volume) discretization segments a whole domain into several cells. For each cell, the numerical solver solves the governing equations mentioned below in equations 1-2 and calculates all the necessary physical variables. Discretization is crucial and selective, depending on the type of problem. The present study uses the SnappyHexMesh tool of the OpenFOAM platform to discretize the computational domain. This tool initiates with a basic Cartesian block mesh of 8 mm cell size throughout the entire domain. It provides a body-fitted mesh as an output with layers of refined cells near the domain boundaries. These cells are refined to such an extent that they obey the $y^+$ constraints at the domain boundaries (near the wall $y^+ \sim 1$). For the regions away from the wall, this study applies an efficient mesh refinement tool, AMR, to refine the mesh far away from the wall to achieve the target LES quality index.

We investigate the transient development of column vortex with time. The flow changes over time as it moves across various areas of the vast computational domain. In this specific study, the areas of interest dynamically shift throughout time. In this transient problem, static refinement, i.e. a one-time refinement at the beginning of the computation leads to computational inefficiencies and unnecessary high computational expense. In light of this, we use the Adaptive Mesh Refinement (AMR) tool. Vorticity magnitude represents the gradient in velocity data, which, in turn, represents all of the critical evolutions in the flow field, especially in incompressible flows such as this case. Hence, the present study uses vorticity magnitude as a decisive parameter for the local mesh refinement. Regions with vorticity magnitude falling within a range of (50 and 800) s$^{-1}$ are refined further in every interval of four time steps while the mesh in regions with vorticity magnitude outside the former range gets unrefined. The 'LES Quality Index' subsection discusses the success of such dynamic refinement in detail. Figures 6-7 present the mesh on the H1 plane and how the AMR tool refines only the regions of interest to resolve the flow physics.

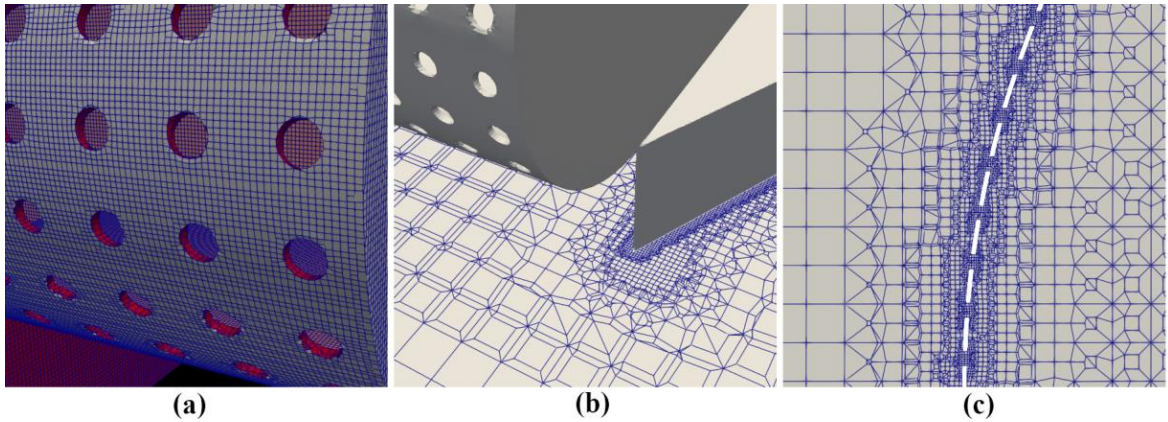

Figure 6: (a,b,c): Some snapshots of the discretized domain using SnappyHexMesh

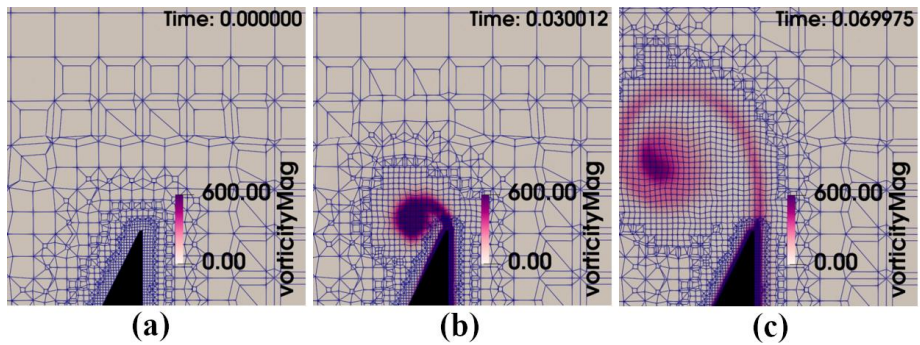

Figure 7: Adaptive mesh refinement (AMR) meshing only the regions of interest. (a, b, c) shows different snapshots of the mesh topology

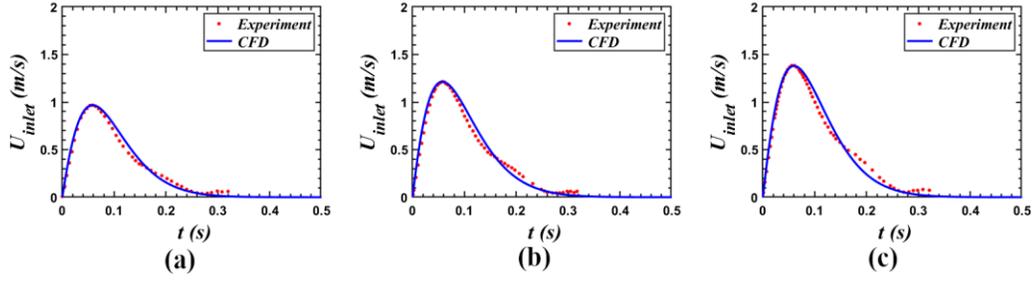

Figure 8: Flow velocity at nozzle inlet, (a) 12V case, (b) 16V case, (c) 20V case

**B. Initial and Boundary Conditions**

Initially, the air is stagnant and the air velocity is zero vector (in all Cartesian directions). Additionally, the pressure at the beginning of the problem (i.e. at t=0) is the same as the atmosphere, 101.325 kPa. Boundary conditions are crucial in any numerical simulation, as these conditions describe the signal provided to the system from the surroundings. The dynamics of a system respond to these signals following the governing physics behind it. The problem statement has already been discussed in the previous section. From a mathematical point of view, these boundary conditions are essential while solving the governing equations mentioned in the coming subsection, 'Flow Solver'. Figure 2 depicts a schematic view of the geometry of the model. The piston drives the air inside the driving compartment and then into the nozzle. The driving piston voltage affects the piston velocity in different Reynolds number scenarios. The driving compartment is not included in the entire computational region to simplify things and save valuable calculating time. Rather, we have regarded the nozzle-connected entrance to the experimental compartment as the computational domain's input, coming from the driving chamber. A white pointer that is fastened to the piston is used to record the piston's velocity. At 200 Hz, a high-speed camera with a 12-bit CCD, 8 MP, and a 100 mm lens records the piston displacement. Using straightforward first-order numerical derivatives, the piston's time-displacement plot is used to calculate its velocity.

The flow velocity at this inlet is calculated from the recorded piston velocity using a simple mass conservation equation for incompressible flow. Figure 8 presents the flow velocity at the inlet varying with time for three different Reynolds number cases (12V, 16V, and 20V respectively). Transient flow velocity at the domain inlet is implemented using a polynomial fit to the experimental data of nozzle inlet velocity (refer Figure 8). The pressure at the inlet boundary is Neumann type and has a zero gradient value. As stated earlier, all outlet boundaries are open to the atmosphere. Hence, all of the outlet boundaries are set to total pressure conditions. This total pressure is constant atmospheric pressure, i.e. 101.325 kPa. Walls in the model geometry behave as no slip and no penetration boundaries. The pressure at these boundaries is set to zero gradients since there is no airflow across these wall boundaries. Figure 4 portrays all domain boundaries in a schematic representation for a clear understanding of the reader.

**C. Flow Solver**

A numerical solver solves the governing equations in each discretized cell (node) to calculate flow variables (pressure, velocity) at every element and node of the domain. The present study uses the Navier Stokes mass and momentum conservation equations (refer to equations 1-2).

$$\nabla \cdot U = 0 \tag{1}$$

$$\frac{\partial U}{\partial t} + (U \cdot \nabla) U = -\frac{1}{\rho} \nabla P + \nu \nabla^2 U \tag{2}$$

where U and P are velocity vector and pressure, respectively. The equations mentioned above are presented in vector form. $\rho, \nu$ are fluid density and kinematic viscosity. The present study deals with very low Re and incompressible flow of air. Hence, the equations assume density and other fluid properties are constant throughout the domain and time range.

This study applies the PIMPLE (Pressure Implicit with Splitting of Operator) algorithm to solve the present transient problem. This algorithm converges the flow variables in each time iteration using inner loops. The divergence and gradient terms are calculated using Gauss integration of linearly interpolated values on cell faces. The Crank Nicolson scheme is used for time marching. Hence, all of the space and time discretization are second-order accurate.

**D. LES Quality Index**

This article numerically investigates the transient development of a finite column vortex. We use LES (Large Eddy Simulation) to resolve the turbulent structures of various scales. LES resolves large-scale structures and models the small eddies using explicit filtering techniques. The present study models the small eddies using the WALE[47] (Wall Adaptive Local Eddy-viscosity) model. The mesh in the region of interest is refined (using AMR) so that the cell size captures most of the large-scale turbulent flow structures and models the rest (smaller than the filter width).

The index of LES quality measures the resolution of turbulent flow structures, which justifies the quality of the mesh used for LES based simulation. Celik et al.[48] provide a measure of the LES quality index regarding kinematic and effective viscosity (refer to equation 3).

$$LES_{IQ} = \frac{1}{1 + \alpha_v \cdot \left(\frac{\nu_t}{\nu}\right)^n} \tag{3}$$

For a decent LES quality ($LES_{IQ}$ above 80%) and a DNS quality ($LES_{IQ}$ above 95%), $\alpha_v$ and $n$ are taken as 0.05 and 0.53 respectively[48,49]. And where $\nu_t$ and $\nu$ are turbulent and kinematic viscosity, respectively. Figure 9 portrays the index of LES quality obtained using the above equation 3. The index falls within an overall range of (0.92 to 1.0). Specifically, in the region of present interest, the quality index is reportedly larger than 0.94. According to Celik et al.[48], such a good quality of LES is equivalent to DNS (Direct Numerical Simulation) results.

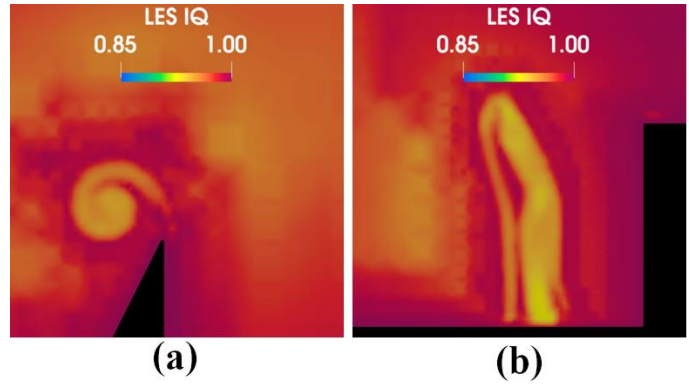

Figure 9: (a) LES IQ in plane H1 (y=0.025 m plane); (b) LES IQ in V1 (x=0.25 m) plane for 20V case.

**E. Data Validation**

We compare our numerical results with our experimental data before moving on to any additional analysis and conclusions. The dynamic evolution of the vortex column is the primary focus of the current effort. The growth and advancement of the vortex column in space and time are of great interest. We investigate the vortex core strength evolution and the trajectory of the core experimentally on the H1 plane. Rotationality in the core region creates a local pressure depression inside the vortex core. The pressure is the minimum, and vorticity is the maximum at the center of the core. In this work, we utilize the minimum pressure criterion to track the travel of the vortex core over time. Figure 10 presents the locus of the vortex core in the domain frame for clear understanding. Figure 10 (a) shows the pressure depression contour inside the vortex core. Figure 10 (b-f) compares the vortex trajectories of experimental and numerical results for five different cases based on five different piston voltages. Circulation is the line integral of the velocity field along a closed curve. A circle is carefully chosen to enclose the vortex core inside it and exclude the outer spiral part of the vortex sheet that is not considered a vortex core. The circle centers on the core center, i.e. the point of pressure minima in the vertical region. The radius is carefully decided based on the extent of the vortex core. Each time the vortex moves its position, the circle and its radius change to a set of new values. The trajectory in the present study falls within the 5% error band w.r.t the experimental observations.

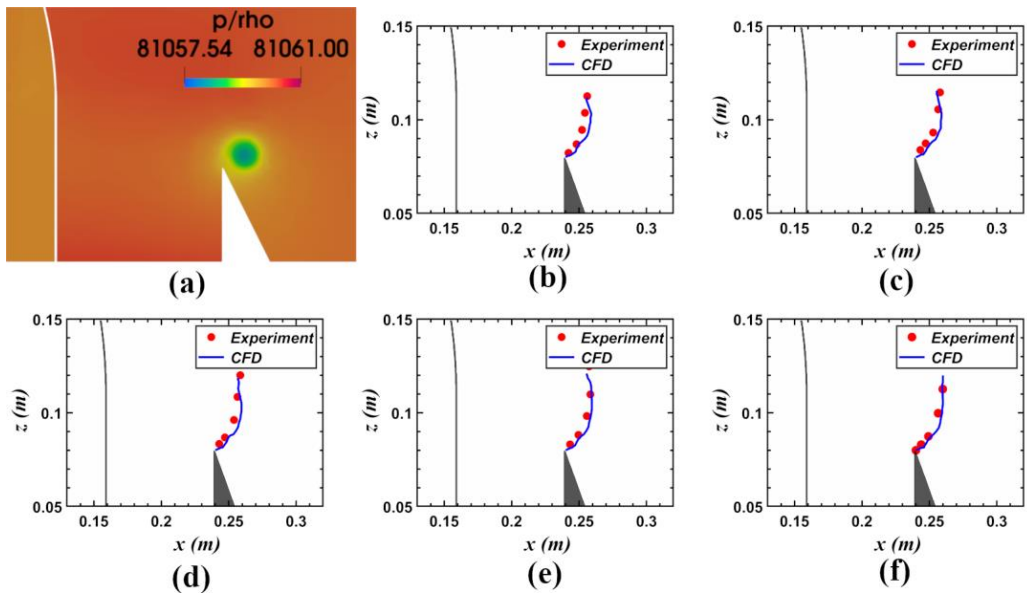

Figure 10: (a) Pressure contour inside the vortex core;
Validation of numerical data for vortex trajectory on H1 plane: (b) 12V, (c) 14V, (d) 16V, (e) 18V, (f) 20V cases.

Vortex strength (circulation) inside this circle keeps increasing as the vortex travels downstream in the bulk flow direction. Due to continuous flow separation, the rolled-up vortex sheet keeps dumping its strength into the vortex core. Vortex core strength monotonically increases as the inlet flow rate keeps on increasing. We track the evolution of the vortex up to only 0.11s of flow time. Figure 11 compares vortex core strength in numerical simulation with the experimental observations. Figure 11(a) presents the choice of a circle around the vortex core to estimate the circulation data. Figure 11 (b-f) portrays the numerical validation for vortex strength for five cases based on five piston voltages. The numerical results fall within the 10% error bound of the experimental results.

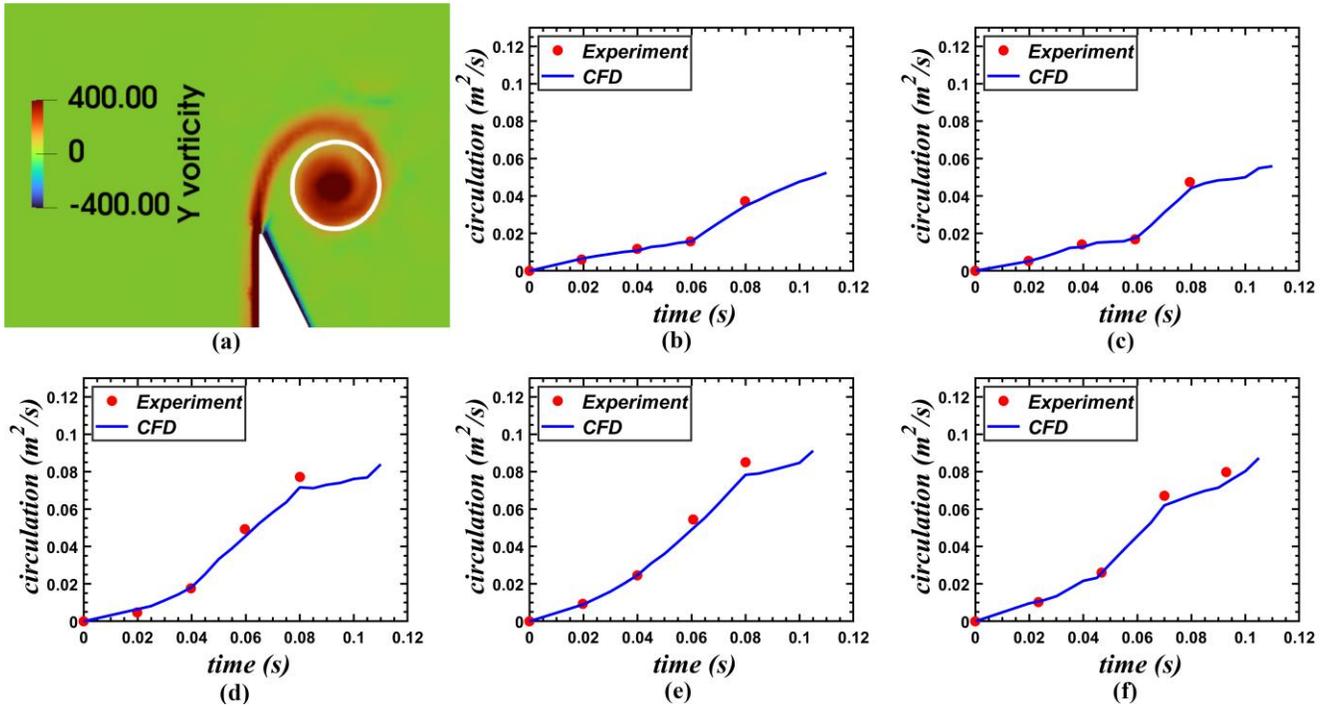

Figure 11: (a) Choosing a circle enclosing the vortex core;

Numerical validation of vortex core strength on H1 plane: (b) 12V, (c) 14V, (d) 16V, (e) 18V, (f) 20V cases.

## V. RESULTS AND DISCUSSION

Concerning the cross-sectional evolution of the column vortex, we mainly study the trajectory and evolution of the vortex strength. The ultimate goal of this effort is to determine the column vortex's three-dimensionality along its length. The mechanics of vorticity transport, vortex column buckling, and torsion are revealed during the temporal evolution of the vortex. Vortex lines and streamlines are evidence of such deformation and evolution of the vortex. This article thoroughly discusses the evolution and advancement of the vortex core line. The vortex column buckles and twists along the length. The degree of these kinds of deformation varies with time and the length of the vortex. The following subsections discuss the buckling of the vortex column and its probable reasons. Buckling in the vortex column is co-related with several other essential dynamics, such as meridional swirl (axial flow). Meanwhile, the perturbations originating from the complex model structure and interference of several flow features induce bending instabilities in the vortex column. These perturbations are spontaneous and do not follow any particular user-defined frequency or wavenumber. The primary focus of this work is on how these spontaneous perturbations lead to flow instabilities and how these instabilities affect the dynamics and shape of the vortex core. A thorough examination of this three-dimensional cumulative behavior, together with our encouraging experimental findings, aids in our comprehension of the real-world applications in natural column vortices like dust devils and tornadoes. In the upcoming subsections, a three-dimensional representation of the vortex column clarifies these concerns.

The centroid of the vortex core displaces from the primary axis of the vortex column due to long wave bending instabilities. The circular core takes an elliptic shape with time. The following subsections explain the shape change and splitting of the vortex column. A quantitative study on extraction and geometrical analysis of vortex core lines reveals several conclusions for all cases (for different piston voltages). The trajectory of the vortex core at different cross-sectional planes is different owing to the bending instability, as discussed earlier. The flow separation in the transverse plane (on the diverging surface at the top wall of the nozzle) is responsible for the origin of meridional flow in the vortex core. On the other hand, the meridional swirl inside the vortex core is responsible for the axial vorticity transport and core size variation of the vortex at different lengths and variation of axial flow in different lengths of the vortex column that leads to column buckling. The whole discussion is divided into the following four subsections. The following subsections also elaborate on the spatial and temporal variation of the meridional flow as well as the helical shape of the vortex column. This article considers five different case setups (as discussed earlier) and investigates the evolution of three-dimensionality in the column vortex with time, space, and variation of inertia forces or Reynolds number (Re).

**A. Meridional Flow and Vortex Lines**

As mentioned by Melander *et al.*[50], cross flow in the meridional plane along the axis of the vortex transports the vorticity along the length of the vortex. This leads to axial variation of vorticity and the core size of the vortex. Meridional flow creates a significant transverse component (rotation in length-wise plane, meridional swirl) of vorticity along with the axial component (cross-sectional plane rotation, sagittal swirl). The non-zero transverse component of vorticity with dominant cross-sectional vorticity results in the helical shape of the vortex lines. Vortex lines are almost parallel to the column structure of the vortex as the sagittal swirl is dominant over the meridional swirl. Regions where the meridional swirl is significantly present impose a twist in the local vortex lines. Figure 12 portrays the shape of the vortex lines inside the vortex column. It is very evident from the figure that these vortex lines do not have the same degree of helix everywhere. Structures of vortex lines imply the non-uniform helical nature of the vortex lines inside the column vortex region. Observations also suggest that the vortex lines nearer to the core line experience a less helical nature than those nearer the column vortex perimeter. The sagittal swirl is stronger in the vortex's central region, which is responsible for such kind of behavior of these vortex lines. The vortex in the transverse plane grows with time and travels downstream in space following the vortex roll-up trajectories[14]. This vortical flow and swirl in the transverse plane affect the meridional flows in the vortex column. This meridional flow evolves in different regions of the vortex column, and the column travels downstream with the nozzle jet flow with time. Effected regions of the vortex column by the meridional flow keep changing with time, too.

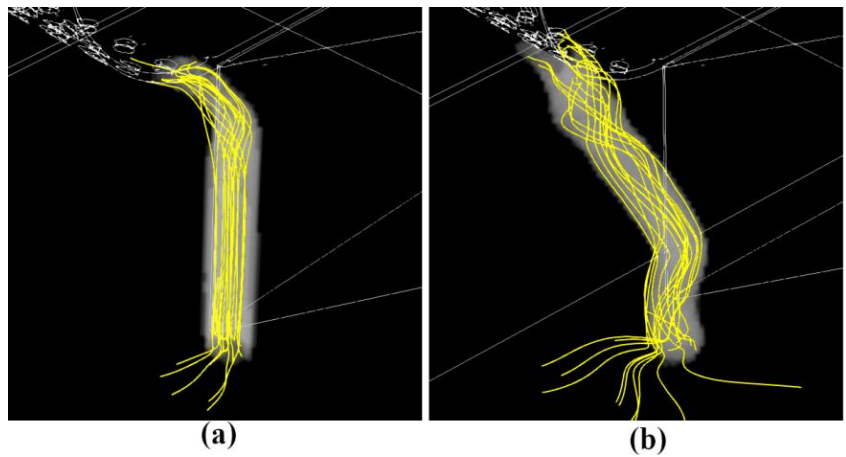

Figure 12: Three-dimensional vortex lines at (a) 0.03s and (b) 0.07s in 20 V case

As time passes, the outflow from the nozzle diverges from the face C while the flow velocity at the nozzle inlet increases. The inertia force increases with time, and the flow overcomes the threshold of the flow separation (zero wall shear stress). Flow separation in the transverse plane is slower than the primary flow separation (responsible for the sagittal swirl in the vortex column) from the edge of the triangular wedge. This phase lag plays an important role in the buckling of the vortex column. Flow separation in the transverse plane creates a vortex that detaches from the face B and contributes to the axial flow in the vortex column. Figure 13 presents the flow separation process in the transverse plane for three different times (0.05s, 0.07s, and 0.09s) for the 20V (piston voltage is 20V) case. This figure considers a slice-plane at x=235 mm position that cuts the nozzle and faces C. Initially, the flow on face C remains attached for some time. As the flow velocity at the inlet and corresponding inertia force rise, the flow separates from the face C, leaving behind a swirl on the transverse plane. The second row of this figure presents the isometric view of the columnar vortex at three different times. With time, as the flow separates from the face C, the upper end of the columnar vortex detaches from the wall too. This flow separation in the transverse plane induces a length-wise axial flow (meridional swirl) inside the objective column vortex. This length-wise flow transports vorticity along with it.

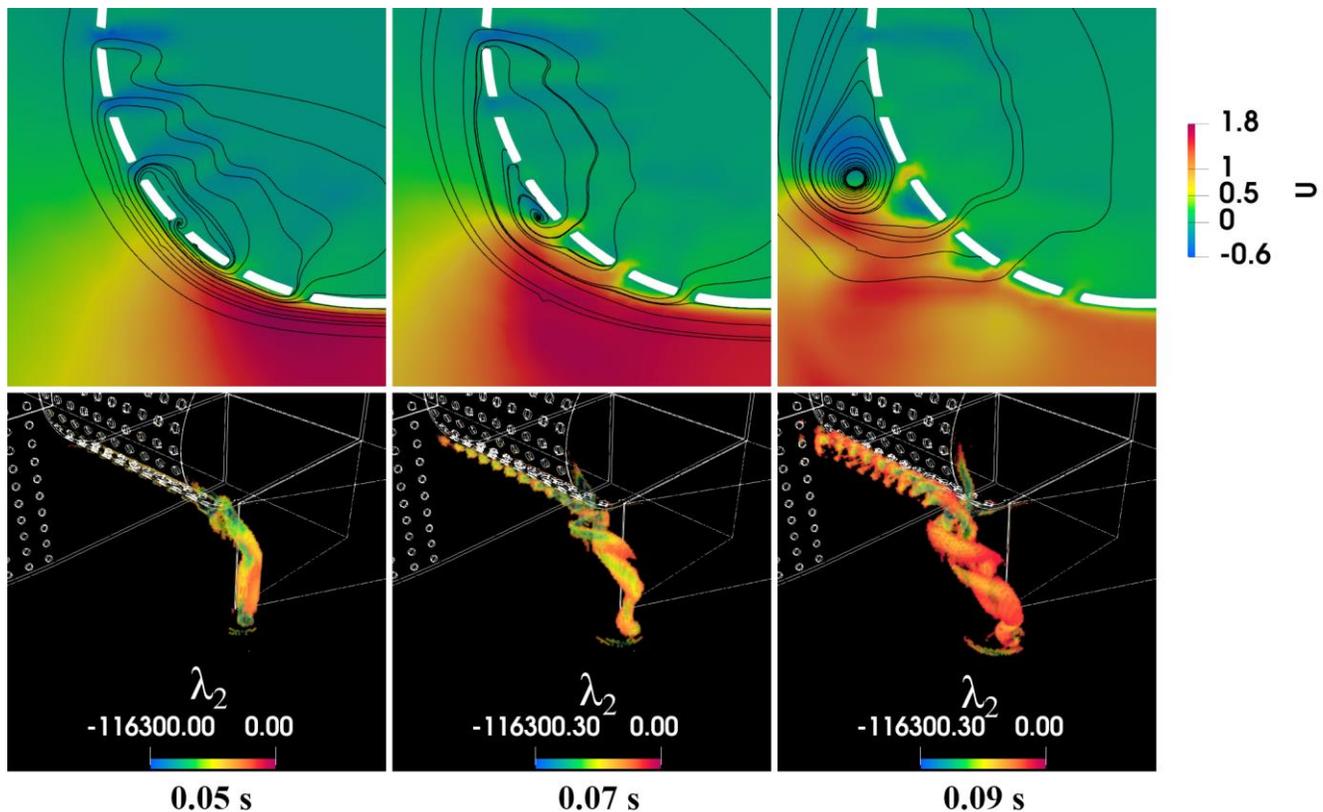

Figure 13: Flow separation and column buckling on the transverse plane (x=235 mm) at 0.05s, 0.07s, and 0.09s in the 20V case. Upper Row: Streamlines highlighting the flow separation process while the contour represents the Z-directional flow velocity (U). Lower Row: Isometric view of vortex core column presenting the three-dimensional overview of the separation process.

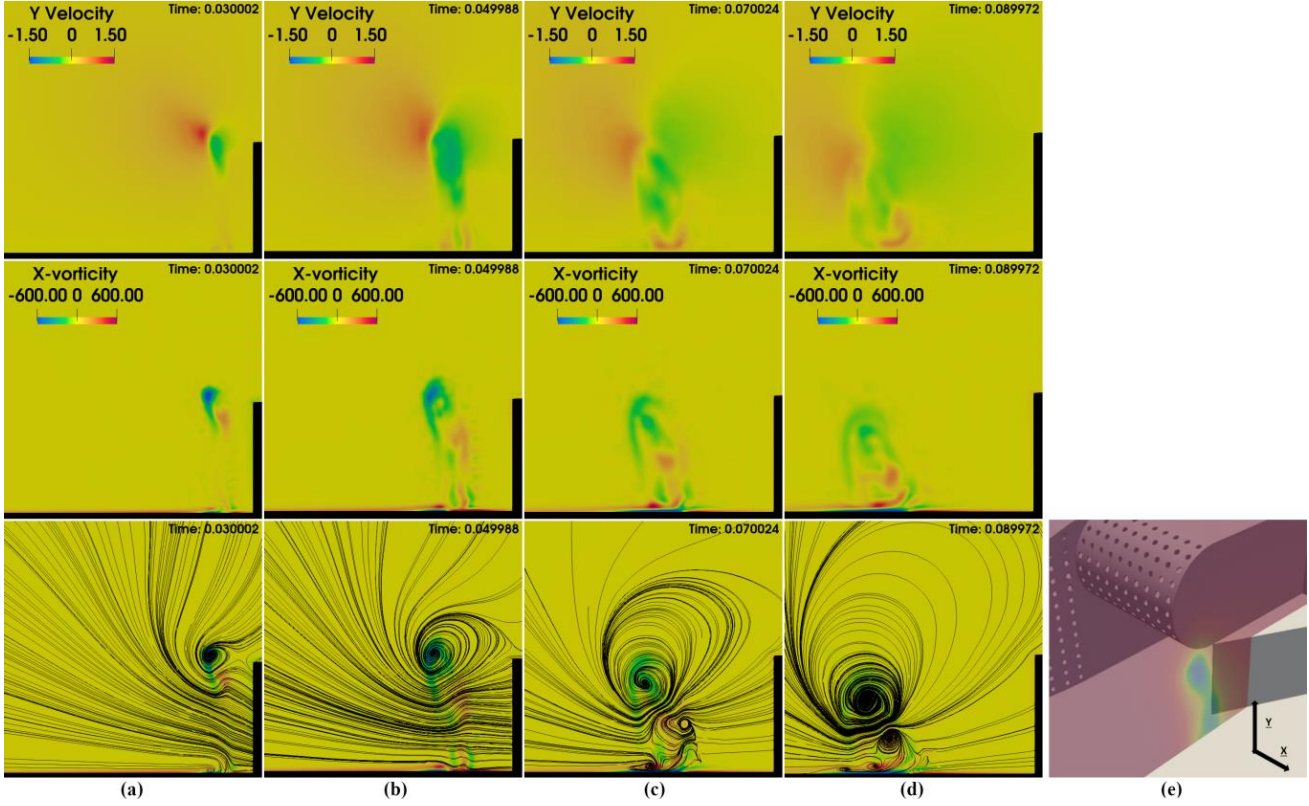

Figure 14: Y-velocity (length-wise velocity), X-Plane vorticity (meridional vorticity), surface streamlines on the mid-vortex meridional plane. At four different times: (a) 0.03 s, (b) 0.05 s, (c) 0.07 s, (d) 0.09 s. (e) the meridional plane is chosen at any time instant and it passes through the maximum vorticity point inside the vortex core on the H1 plane.

Figure 14 portrays the meridional plane of the column vortex at four different times in the 20V case. These slice planes are meridional planes (X-normal planes), passing through the maximum $\lambda_2$ [51] (maximum sagittal swirl) position on the H1 plane and inside the vortex core. $\lambda_2$ is defined by Jeong *et al.* [51] as the second biggest eigen value of $(\Omega^2 + S^2)$, where, $\Omega$ is the vorticity tensor (antisymmetric part of velocity gradient tensor) and S is the Strain tensor (symmetric part of velocity gradient tensor). $\lambda_2$ signifies the local depression in pressure field and the strength of the rotational flow in the same locality.

The columns in Figure 14 correspond to meridional vortex mid-planes (X-planes) at 0.03, 0.05, 0.07, and 0.09 seconds, respectively (from left to right). In each time column, the first row represents the y-velocity contour, i.e. the axial flow velocity contour. The second and the third rows represent vorticity in the meridional plane, i.e. x-vorticity and surface streamlines, respectively.

Meridional swirl gradually travels downwards and interacts with the boundary layer at the bottom end of the vortex in the vicinity of the bottom wall. The interaction between the meridional swirl and bottom wall boundary layer perturbs the boundary layer, and a chaotic motion of counter-rotating vortices is observed in Figure 14. The interaction of the meridional swirl initially causes boundary layer separation and local flow reversal at the bottom wall. Transient development of these counter-rotating vortices from the interaction of meridional swirl and bottom boundary layer is evident in Figure 14. Near the bottom wall, the meridional swirl plays a dominant role due to a lack of inertia forces and dominance of the viscous effects. This results in a twist in the vortex lines near the bottom boundary of the vortex. This interaction further approaches far more chaotic flow behavior. However, this article investigates the vortex column before any such chaotic flow destiny.

**B. Bucking of the Vortex Column**

The flow separation in the transverse plane is slightly delayed than in the H1 plane (primary flow separation) as the solid surface on face C diverges more slowly than face A in the triangular wedge structure. The holes on face C contribute to this process, too. The flow remains attached to face C for a longer time while the column vortex travels downstream (+Z direction). The top end of the vortex remains attached to face C while the rest of the vortex body flows downstream. This causes the buckling of the column.

The degree of buckling varies in space and time in the column vortex. Buckling of vortex column comments regarding the inertia and viscous effects, effects of delayed flow separation, and so on. The core line of the vortex column provides a qualitative overview of the degree of buckling in different lengths of the vortex column. The vortex behaves according to the Rankine vortex model in the present study. In the Rankine vortex model, the core can be identified by a local pressure depression and rise in local vorticity or $\lambda_2$ value. The present study finds the centroid of pressure minima (in case there are split cores) to locate the vortex core center. In the case of a single vortex core, this inspection returns minimum pressure positions. Such a two-dimensional investigation locates the vortex core center in each Y-plane, i.e. in each cross-sectional plane of the vortex column. Altogether, these points define the vortex core line. Numerical extraction of the vortex core lines brings sharp turns in the segments of this line. The exponential smoothing technique smooths out such sharp turns, which further helps extract important geometric features such as local curvature and torsion.

Figure 15 clears the idea of buckling of the vortex column. The first row of this figure presents the front (X-Y plane), top (X-Z plane), side (Y-Z plane), and isometric (three-dimensional) view of the vortex core line. The core line is smoothed out to eliminate all small numerical fluctuations (due to discrete grid points). The first row corresponds to t=0.07s time instant of the 20V case. The second row represents four instances of the same case to show the temporal development of the buckling of the vortex column. The buckling phenomenon is reported till the attachment of flow in the transverse plane, as discussed in the earlier subsection. After the separation of the top end of the vortex column from the face C, the buckling effect on the column vanishes.

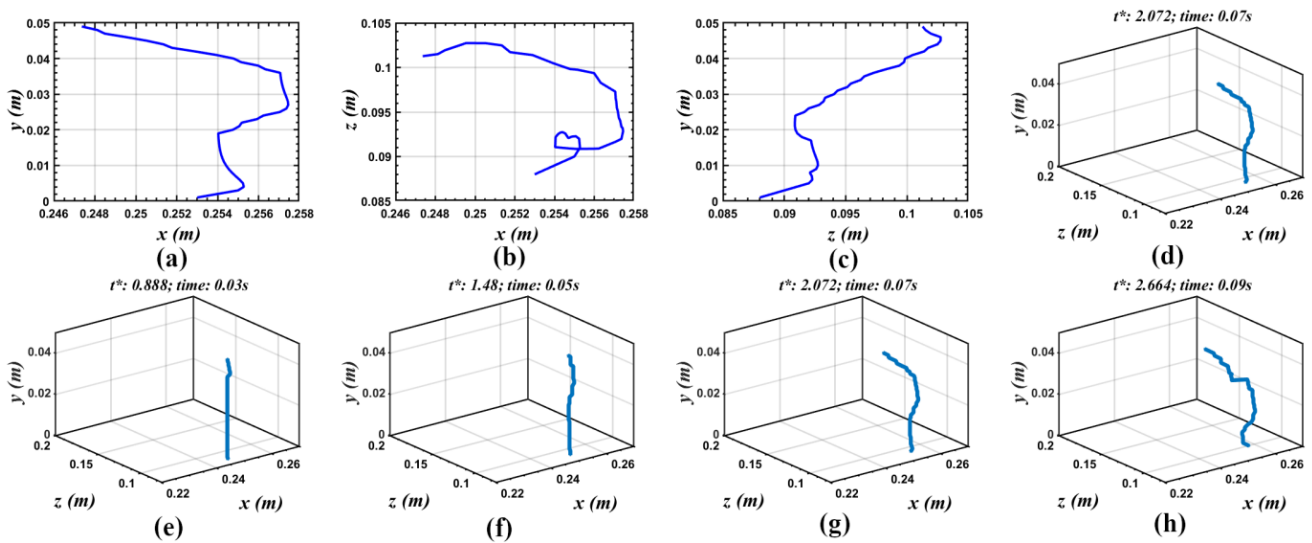

Figure 15: Front (a), top (b), side (c), and isometric (d) view of the vortex core-line for 20V case at t=0.07s. Isometric view of vortex core-lines at t=0.03s (e), 0.05s (f), 0.07s (g), and 0.09s (h) for 20V case.

## C. Bending of the Vortex Column

This section presents a three-dimensional vortex column with cumulative vortex bending and buckling effects. Bending instabilities in vortex columns have been studied from numerical and theoretical perspectives. As mentioned earlier, the columnar vortex in the present study resembles the Rankine vortex model. Rankine vortex combines solid-body rotation at the vortex core and irrotational potential flow around it. In the present study, a belt of non-swirling straining region surrounds the vortex core at its center. Kelvin[15] discovered the instability due to minor disturbances at the core. Chandrasekhar[16] and Krishnamoorthy[17] demonstrated the contribution of axial flow to instability in the vortex column. Axial velocity plays a vital role in temporal and spatial instability in the vortex column. It is also important to comment regarding the transition between absolute and convective instabilities. Loiseleux *et al.*[23] related this phenomenon to the vortex breakdown process. An abrupt jump in axial velocity across the vortex core surface leads to short- and some-long-wave temporal instabilities. This kind of instability is popularly known as Kelvin-Helmholtz instability. Perturbations in the vortex core lead to different modes of instabilities, such as axisymmetric (m=0), spiral (m=±1), and higher modes. The behavior of the unstable vortex core line in these modes is a function of wavenumber (k), Swirl number (S=1/Rossby number), and temporal frequency ($\omega$). Loiseleux, Wu *et al.*[52], Arendt *et al.*[53] have provided numerically and theoretically investigated correlations among these parameters that play crucial roles in the instability analysis of vortex cores.

The steep gradient in axial flow velocity in meridional swirl contour is evident in Figure 14. However, the scale of the axial velocity is very small (in order of $10^{-1}$). Meanwhile, the sagittal swirl is comparatively more powerful. The local swirl parameter ($\frac{\Omega R}{\Delta W}$, where $\Omega$ is the local angular velocity, R is the radius, and $\Delta W$ is the scale of jump in axial velocity) is very high, and the Rossby number is relatively low. Low values of Rossby number indicate the dominance of Coriolis force over the local inertia force of the vortex, which is quite evident from the small scale of the axial flow. Careful observations of the core line suggest that the wavenumber is very low. It is evident from Figure 16 that the wavelength varies in different lengths of the vortex. Moreover, from Figures 15 and 16, it is also evident that the wavenumber keeps changing locally in time as the shape and torsion of the core line keep varying over the length of the column vortex and with time (refer to Figure 17). Overall observations suggest that the wavenumber of the vortex column bending is ~2/H mm$^{-1}$ for the present case (H is the

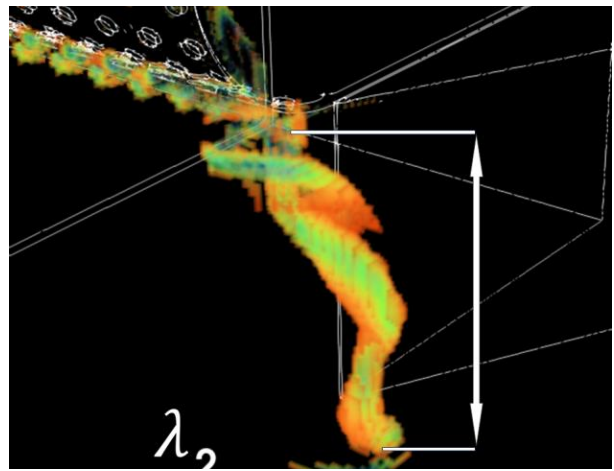
25

Figure 16: Long wave vortex column bending. Wavelength, L ~25 mm, i.e., half the column height.

height of the column vortex, 50 mm in the present case). The vortex column takes its entire length to complete one revolution of the helix (refer to Figure 16).

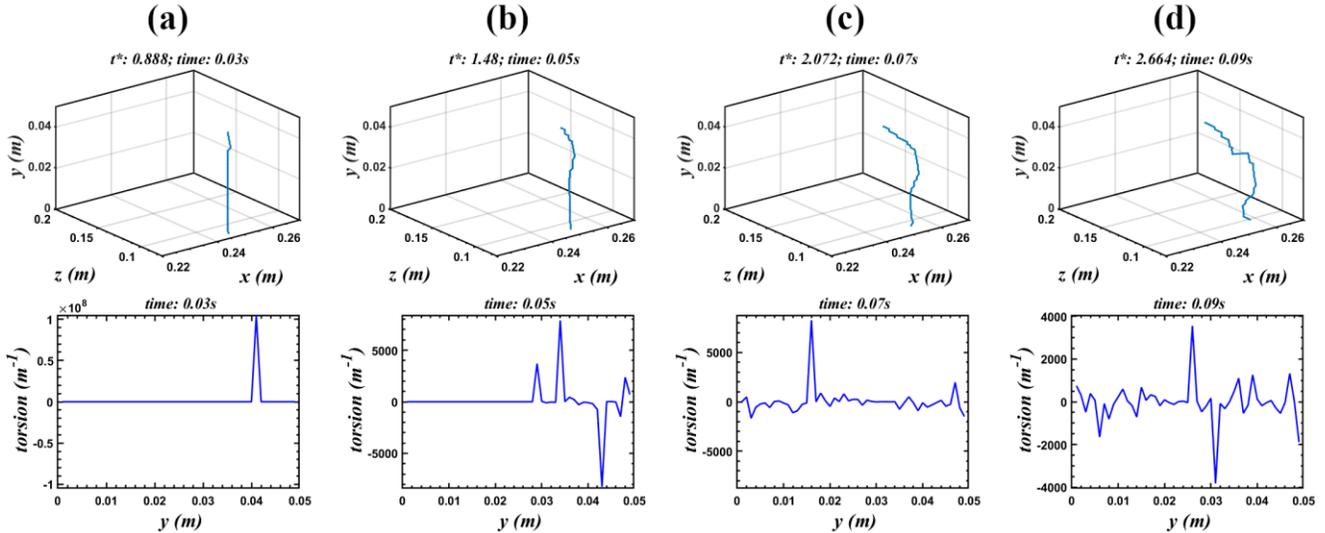

Figure 17: Torsion of Vortex core-line plotted against the length of the core-line for four different time instances for the 20V case: (a) 0.03s, (b) 0.05s, (c) 0.07 s and (d) 0.09s

Another careful observation of the bending of the vortex column tells that in the present study, the vortex column takes the form of the left-handed helix of the wavenumber mentioned above. Only the spiral (m=+1) instability mode with long-wave column bending is observed in the vortex column. The bending phenomenon starts with the formation of an elbow at the mid-length of the vortex column. Arendt et al.[53] presented results regarding spatial instability in the vortex core. Initially, a strained element of the vortex tube bends towards an orthogonal direction and starts rotating around the tube. This gives rise to a pair of left and right-handed helix traveling towards opposite directions along the vortex tube. However, in the present case, only one helix type is observed due to the finite and concise length of the vortex column. The elbow is defined as the meeting point of the inclined part of the buckled column and the straight part.

Perturbations in the present study originate from the complicated model structure mentioned in earlier subsections. This study uses no turbulent intensity at inlet flow while implementing the boundary conditions. The key objective of this study is to investigate the evolution of the vortex column and to conclude with proper explanations that will be helpful for studying and predicting the natural vortices. As the perturbations and turbulence originate from the geometry near the model walls, the WALE model perfectly computes the fluctuating flow fields. Figure 18 presents the extracted core line for the 20V case. Each row represents a time instant (from 0.03s to 0.09s). The variable 's' represents the length along the vortex column starting from the bottom end. The bending phenomenon (spiral mode instability) drifts the core center from the straight axis at all 'y' positions.

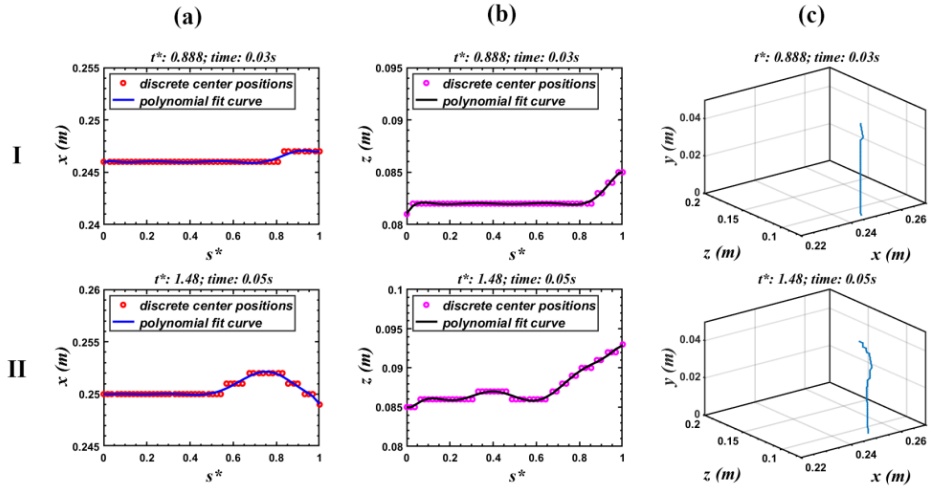

Figure 18: Represents drift of the x-position (column a) and z-position (column b) of the vortex core center with the length of the column. (c) Presents the isometric view of the vortex core line. I) 0.03s, II) 0.05s, III) 0.07s and IV) 0.09s time instants for 20V case.

The first two columns of Figure 18 portray the plots of the x-position and z-position of the coordinates in the vortex core line versus the variable 's'. 'Z' is the direction of the piston-driven flow and also the downstream traveling direction for the column vortex. The 'Y' axis is directed along the height of the column vortex and 'X'-direction is the transverse direction. These plots at different times provide a proper idea of the temporal evolution of the bending effects on the vortex core. The discrete points are fit against a ninth-order polynomial for proper visualization. The unstable wavy features are prominent in this figure. The wave and fluctuations start from the top end of the vortex column at the joint of faces B and C. Gradually, this wave travels downwards along the length of the vortex till the bottom end, and the axial flow interacts with the bottom wall boundary layer. Local curvatures and bending in the vortex core line appear due to the attached flow until 0.07s on face C. After the flow separation in the transverse plane, the buckling effect vanishes gradually, and the core line achieves linearity in z-positions.

The smoothed-out data of the vortex core line provides the idea of torsion in the vortex column at different lengths of the column. Torsion quantifies the degree of twist and the effect of the bending mode on the column. The top end of the vortex column achieves high torsion at the initial time. The torsion effect travels gradually downstream with time (Figure 17). Figure 19 provides the isometric views of the vortex column for four different time instances and five different case setups (based on piston voltages). These figures throw more light on this topic. The vortex core gets split into two daughter cores (discussed in the coming subsection). This phenomenon also begins at the top end of the vortex column. The meridional flow also originates from the top end of the vortex (from the face C). In the following section, further explanations regarding the vortex core splitting correlate with the whole dynamics of the vortex column in this study.

**D. Splitting of the Vortex Core**

As mentioned earlier, all of the phenomena mentioned above start from the top end of the vortex and gradually keep affecting the lower portions of the vortex column. The down travel of such phenomena is not any kind of diffusion process. Instead, these are types of interactive phenomena.

As the face C diverges from the flow direction, the flow on face C remains attached initially, and then the flow diverges from face C. On the other hand, the face E is straight. The rate of flow divergence is different on this face than on face C. Several literature show that faces with holes without any imposed blowing (suction) phenomena catalyze the flow separation process more than simple continuous straight surfaces. The present case witnesses such a scenario together. The flow separation on face C is faster than on face E. Initially, the vortex remains attached to two faces, C and E, respectively. Gradually, as the lag in the flow separation process appears distinct with time, the vortex core gets bifurcated into two daughter cores of different strengths and with the same sense of rotation. Meanwhile, the vortex core travels downstream while revolving around the local axis. This revolution of the mother vortex core and its daughter cores results in the braided structure of the daughter cores. One end of these daughter cores is attached to the solid boundary (face E) and secondary vortices in the transverse plane (separated flow on face C), while on the other end, these daughter vortices meet the mother vortex column.

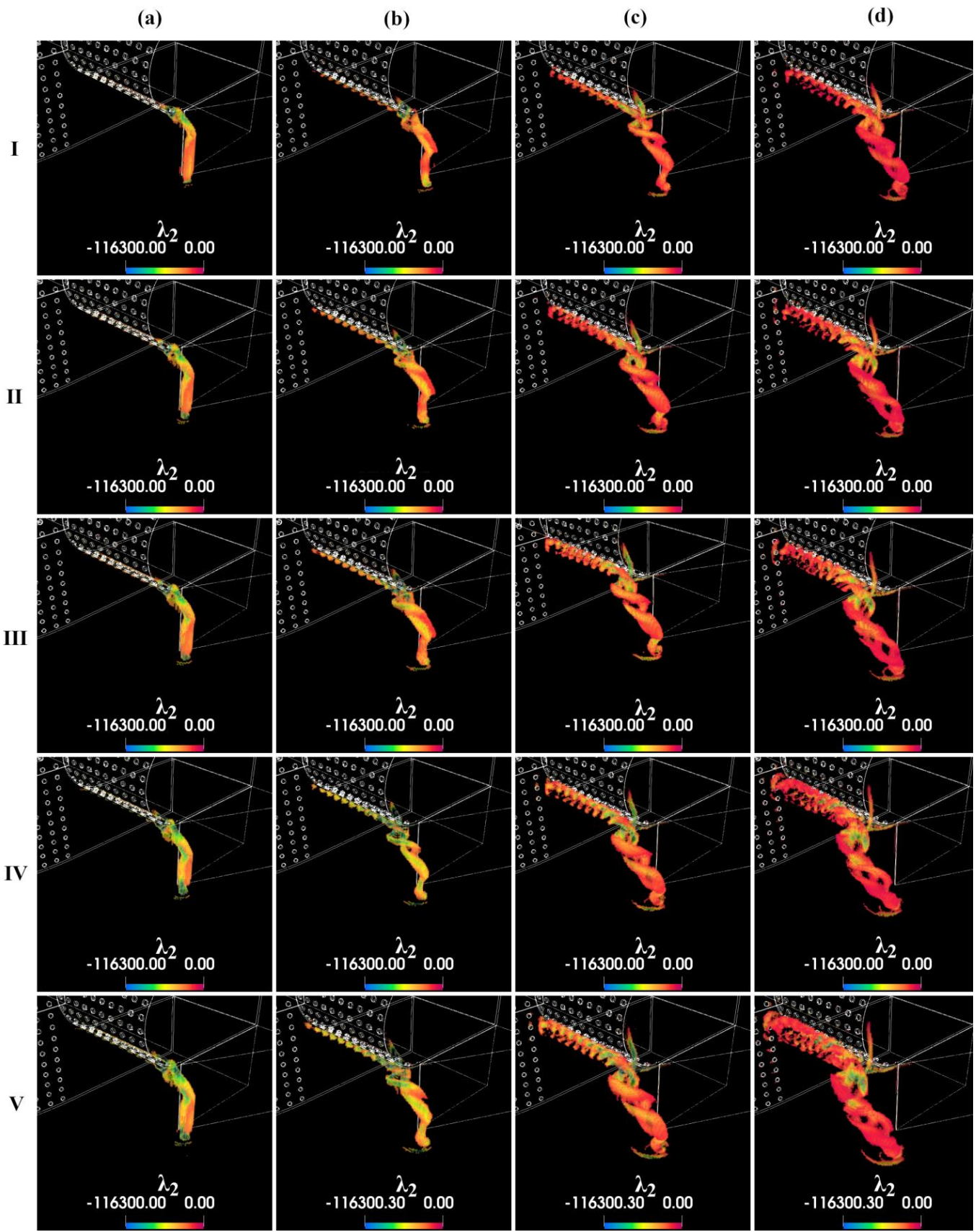

Figure 19: Isometric view of the splitting of vortex column into two helical daughter cores. The columns represent four different time instants: (a) 0.05, (b) 0.07, (c) 0.09, (d) 0.11s. Rows represent 12V (I), 14V (II), 16V (III), 18V (IV) and 20V (V) cases.

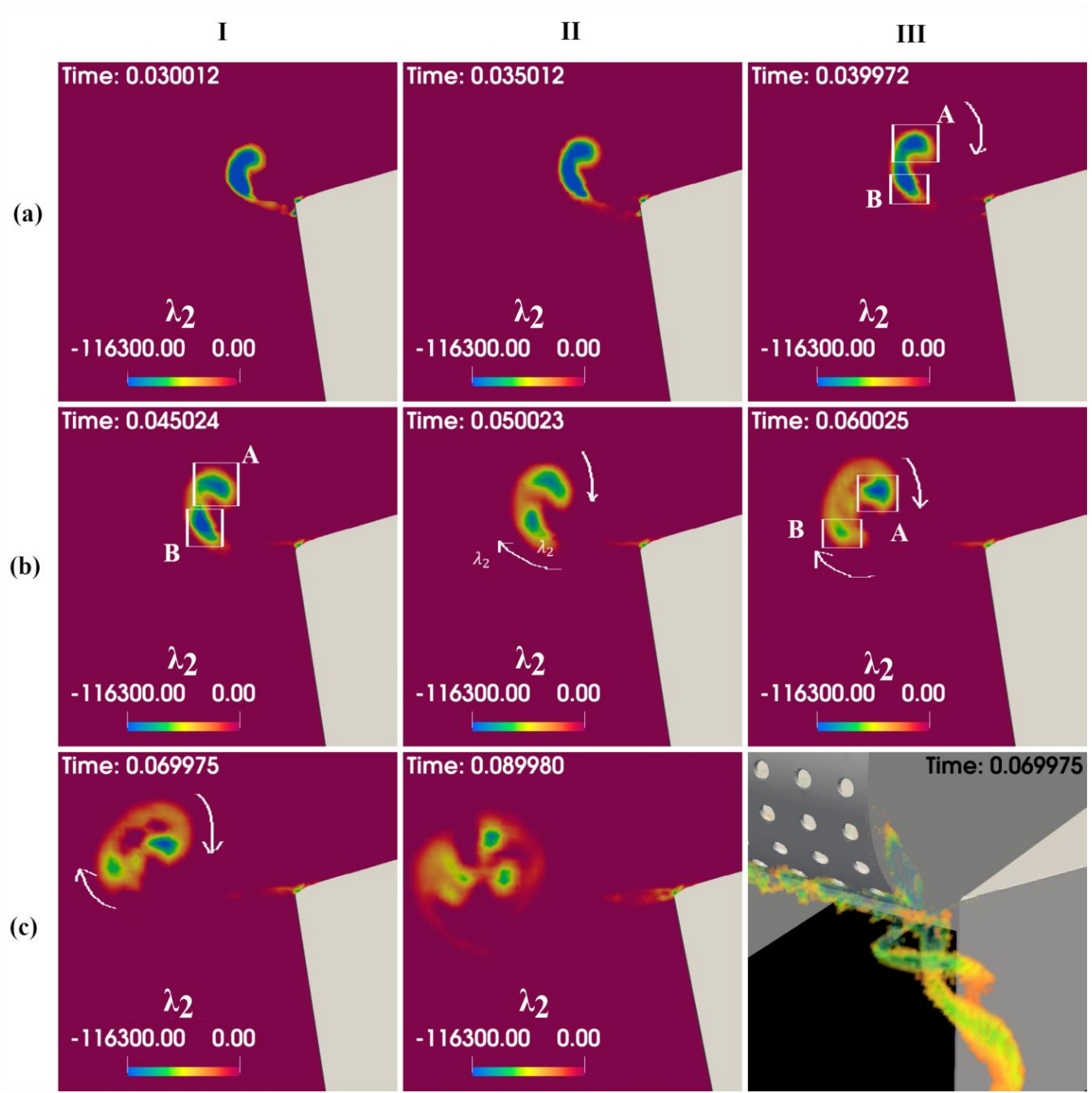

Figure 20: I, II and II-(a): Splitting process of vortex core on a plane P near the top end of the vortex; I, II and III-(b): Daughter vortices A and B split up and start revolving in the CCW direction. III-(c) : Isometric three-dimensional view of the mother and daughter vortices in braided form near the top end of the vortex column.

Both of the daughter cores take a left-handed helix just like their mother. With time, the vortex core keeps revolving around the local axis, and the splitting point (where the mother vortex core gets bifurcated into two daughter cores) travels downwards along the length of the mother vortex column. This advancement of splitting phenomena along the vortex column length depends on the inlet flow based Re. For a better understanding of the splitting phenomena, a plane 'P' (normal: [0.707, -0.707, 0] in meters) is considered near the sharp edge, i.e. face B and the top end of the vortex (refer to figure 19.

This plane cuts the vortex column at the bend (just before the vortex column ends in the solid surface at the top end). Figure 20 presents a total of eight different time instances of $\lambda_2$-contour that identifies the vortex core at this bend and one isometric view of the bifurcated vortex column near the top end (refer Figure 20-III-(c)). This position is critical for the present investigation as the vortex column bifurcates from this point. The time instances are mentioned in the contour plots to help the reader to understand the chronology of the snapshots. Figure 20-I-(a) presents a single mother vortex core in the initial phase. This core grows in Figure 20-II-(a) towards bifurcation. Figure 20-III-(a) portrays the onset of the splitting process and the tendency of the two daughter vortices, A and B (say), to separate from each other. Figure 20-I-(b) presents the separated daughter cores (A and B) with a non-vortex region in between. Eventually, they start revolving around each other in the same sense as the mother vortex (Figure 20-II-(b), III-(b), and I-(c)). The sense of rotation in each daughter vortices is also the same as the mother vortex core (counter-clockwise). Initially, after the splitting, the strengths of the vortex cores are on a comparable scale. The daughter core A's strength gradually dominates the other daughter core B as the $\lambda_2$ contours in figure 20 suggest.

Understanding the behavior of the column vortex in specific scenarios is made easier by researching the splitting and strength distribution of the cortical core and its daughter cores. Beyond the two-dimensional vortex core trajectory and strength evolution, more such facts are revealed when viewed from a three-dimensional perspective. Predicting the line vortex as it emerges from the edge of the triangular wedge requires careful consideration of numerical data, detailed analysis, and experimental measurements. A three-dimensional summary of core-splitting, bending, buckling, and other related processes provides opportunities for future forecasting of an ex-situ cyclonic vortex's destiny.

## VI. SUMMARY

The combined effects of core splitting, buckling, and bending in a three-dimensional vortex column are intended to be illustrated in this article. In the laboratory, we create a sophisticated structure model to use flow separation and the Biot-Savart induction process to create a vortex column. Using the PIV data from the experiments, we estimate the trajectory and evolution of the vortex core strength by concentrating on the two-dimensional evolution of the vortex core in the mid-plane of the vortex column. In particular, we analyze the three-dimensionality of the vortex column using the three-dimensional numerical data, and we give a comprehensive description of the vortex column's changing dynamics. The core of a vortex can be identified using several vortex identification techniques such as Q-criterion[54], local pressure depression, $\lambda_2$-criterion[51], etc. This study uses $\lambda_2$-criterion for identification of the vortex core. The complex model structure and the resultant flow behavior induce several kinematics and dynamics in the vortex column. The piston drives the flow inside the model that generates the vortex. Depending on the piston velocity, the Reynolds number (Re) of the inlet flow varies and, the behavior of the vortex core changes. The summary of the investigation is listed below.

- The nozzle ends in three types of faces (face C, face B-A, and face E). These faces diverge from the nozzle flow direction differently. For example, the face B-A, i.e. the triangular wedge, has the fastest flow separation. This triangular wedge creates the objective line vortex. Second comes the face C, which is slowly diverging, but the presence of holes without any suction (blowing) effect catalyzes the flow separation. The slowest flow separation occurs on the face E. The top end of the vortex column remains attached to the junction of faces B, C, D, and E, while the other tail ends on the bottom wall of the model where the viscous effect dominates. Just the opposite of this occurs near the top end of the vortex.

- Flow separation in face C and the diverging flow over this face induces a swirl in the meridional plane of the vortex. Transport of the axial vorticity (sagittal swirl) due to meridional swirl, causes variation of vortex core shape along the length of the column. Axial flow in the vortex column leaves a non-zero transverse-plane vorticity (corresponding to meridional swirl). Sagittal and meridional vorticity brings about a twist in the local vortex lines. The meridional swirl dominates over the sagittal swirl near the core's perimeter, while the opposite is observed near the local vortex center. The rotation scale (regarding vorticity magnitude) in the meridional swirl is much smaller than in the sagittal swirl leaving a high pitch helix in the local vortex lines.

- The variation of axial velocity across the vortex core boundary and the jump brings a bending instability in the vortex column. The vortex core line twists around its buckled axis-line. The column's left-handed helical structure suggests m=+1 mode of bending instability. However, the swirl parameter (S) is very high due to the colossal scale difference between the jump in axial velocity (tiny) and the sagittal swirl (comparatively higher) of the vortex column. The local Rossby number is found to be significantly less. The Coriolis force plays dominantly over the axial inertia force. This investigation concludes the effect of such axial flow causes long-wave bending instabilities in the vortex column with a minimal Rossby number. Buckling and bending of the vortex column result in drifting vortex core at different heights of the vortex column, leaving a varying trajectory of the vortex core at its different lengths.

- The study extracts the core center using the maximum pressure depression or $\lambda_2$ technique[51]. These centers at different column lengths are gathered together to form a line. The curve is smoothened using exponential smoothing to eliminate all discrete and abrupt sharp turns in the graph for the sake of accurate numerical calculation of torsion in the core line at different time instances along the length of the core line.

- The variation in the duration of flow separation on faces C and E causes a bifurcation of the vortex column at its top end near the circular edge between faces C and E. This bifurcation leads to the birth of two daughter vortices of the same strength and the same sense of fluid rotation. As time evolves, these daughter vortices keep winding around one another. They keep growing in length, and the point of splitting of the mother vortex core travels downward along the length of the vortex column (as the column travels downstream). With the increase in inlet flow Re, flow separations occur faster on all surfaces, and the core splitting, meridional swirl, buckling of vortex column, and column bending occur faster for higher Re cases than the lower ones.

Several authors have addressed the effects of bending and buckling of vortex columns and the interaction of co-rotating vortex pairs wound around one another separately in a two-dimensional or three-dimensional perspective. However, these phenomena, like most other phenomena in fluid mechanics, are highly non-linear. Simple superimposition of separate phenomena cannot explain or resolve the interaction of these different phenomena. The present article talks about this missing link, providing an overview of the interference of phenomena mentioned above, for a finite-length columnar vortex and all possible explanations regarding the underlying physics. Experimental observations to track the vortex trajectory and evolution of the vortex strength and numerical data to investigate the three-dimensional dynamics help set a strong ground for further detailed studies regarding the finite-length line vortices in full three-dimensional perspective. A solid foundation for more in-depth research on the prediction of ex-situ cyclonic line vortices in full three-dimensional perspective has been established by experimental observations to monitor the vortex trajectory and evolution of the vortex strength and numerical data to study the three-dimensional dynamics.

## ACKNOWLEDGMENTS

The Science and Engineering Research Board (SERB), India, provides financial support for this research (Grant number CRG/2020/005435). The authors would like to acknowledge the National Supercomputing Mission (NSM) for providing computing resources of "Param Sanganak" at IIT Kanpur, which is implemented by C-DAC and supported by the Ministry of Electronics and Information Technology (MeitY) and the Department of Science and Technology (DST), Government of India. We would like to acknowledge the IIT-K Computer Centre ([www.iitk.ac.in/cc](www.iitk.ac.in/cc)) for providing the resources to perform the computation work. This support is gratefully acknowledged.

## AUTHOR DECLARATIONS

### Conflict of Interest

The authors have no conflict of interest to disclose.

## DATA AVAILABILITY

The data that support the findings of this study are available from the corresponding author upon reasonable request.

## REFERENCES

[1] H. Kaden, "Aufwicklung einer unstabilen Unstetigkeitsfläche," Ingenieur-Archiv **2**(2), 140–168 (1931).

[2] Rosenhead, "The formation of vortices from a surface of discontinuity," Proceedings of the Royal Society of London. Series A, Containing Papers of a Mathematical and Physical Character **134**(823), 170–192 (1931).

[3] F.L. Westwater, *Rolling up of the Surface of Discontinuity behind an Aerofoil of Finite Span* (HM Stationery Office, 1935).

[4] D.W. Moore, "The discrete vortex approximation of a vortex sheet," California Institute of Technology Report AFOSR-1084-69, (1971).

[5] P.G. Saffman, and G.R. Baker, "Vortex interactions," Annu Rev Fluid Mech **11**(1), 95–121 (1979).

[6] J.H.B. Smith, "Improved calculations of leading-edge separation from slender, thin, delta wings," Proc R Soc Lond A Math Phys Sci **306**(1484), 67–90 (1968).

[7] P.T. Fink, and W.K. Soh, "A new approach to roll-up calculations of vortex sheets," Proceedings of the Royal Society of London. A. Mathematical and Physical Sciences **362**(1709), 195–209 (1978).

[8] G.R. Baker, "A test of the method of Fink & Soh for following vortex-sheet motion," J Fluid Mech **100**(01), 209 (1980).

[9] H.W.M. Hoeijmakers, and W. Vaatstra, "A higher order panel method applied to vortex sheet roll-up," AIAA Journal **21**(4), 516–523 (1983).

[10] I. Sugioka, and S.E. Widnall, *A Panel Method Study of Vortex Sheets with Special Emphasis on Sheets of Axisymmetric Geometry* (1985).

[11] G.R. BAKER, and L.D. PHAM, "A comparison of blob methods for vortex sheet roll-up," J Fluid Mech **547**(1), 297 (2006).

[12] J.J.L. Higdon, and C. Pozrikidis, "The self-induced motion of vortex sheets," J Fluid Mech **150**, 203–231 (1985).

[13] S.-I. Sohn, D. Yoon, and W. Hwang, "Long-time simulations of the Kelvin-Helmholtz instability using an adaptive vortex method," Phys Rev E **82**(4), 046711 (2010).

[14] A.C. DeVoria, and K. Mohseni, "Vortex sheet roll-up revisited," J Fluid Mech **855**, 299–321 (2018).

[15] K. Lord, "Vibrations of a columnar vortex," Phil. Mag. **10**, 155–168 (1880).

[16] S. Chandrasekhar, *Hydrodynamic and Hydromagnetic Stability* (Courier Corporation, 2013).

[17] V. Krishnamoorthy, Vortex Breakdown and Measurements of Pressure Fluctuation over Slender Wings. Southampton Ph. D, thesis, 1966.

[18] D. Moore, and P. Saffman, "The motion of a vortex filament with axial flow," Philosophical Transactions of the Royal Society of London. Series A, Mathematical and Physical Sciences **272**(1226), 403–429 (1972).

[19] M.S. Uberoi, C.-Y. Chow, and J.P. Narain, "Stability of Coaxial Rotating Jet and Vortex of Different Densities," Phys Fluids **15**(10), 1718–1727 (1972).

[20] M. Lessen, N. V. Deshpande, and B. Hadji-Ohanes, "Stability of a potential vortex with a non-rotating and rigid-body rotating top-hat jet core," J Fluid Mech **60**(03), 459 (1973).

[21] P.G. Drazin, and W.H. Reid, "Hydrodynamic stability," NASA STI/Recon Technical Report A **82**, 17950 (1981).

[22] P.G. Saffman, *Vortex Dynamics* (Cambridge university press, 1995).

[23] T. Loiseleux, J.M. Chomaz, and P. Huerre, "The effect of swirl on jets and wakes: Linear instability of the Rankine vortex with axial flow," Physics of Fluids **10**(5), 1120–1134 (1998).

[24] P. BILLANT, J.-M. CHOMAZ, and P. HUERRE, "Experimental study of vortex breakdown in swirling jets," J Fluid Mech **376**, 183–219 (1998).

[25] O. Zeman, "The persistence of trailing vortices: A modeling study," Physics of Fluids **7**(1), 135–143 (1995).

[26] S. Wallin, and S.S. Girimaji, "Evolution of an isolated turbulent trailing vortex," AIAA Journal **38**(4), 657–665 (2000).

[27] S.C.C. BAILEY, and S. TAVOULARIS, "Measurements of the velocity field of a wing-tip vortex, wandering in grid turbulence," J Fluid Mech **601**, 281–315 (2008).

[28] D.S. Pradeep, and F. Hussain, "Vortex dynamics of turbulence–coherent structure interaction," Theor Comput Fluid Dyn **24**, 265–282 (2010).

[29] A. Antkowiak, and P. Brancher, "Transient energy growth for the Lamb–Oseen vortex," Physics of Fluids **16**(1), L1–L4 (2004).

[30] A. ANTKOWIAK, and P. BRANCHER, "On vortex rings around vortices: an optimal mechanism," J Fluid Mech **578**, 295–304 (2007).

[31] M. V. Melander, and F. Hussain, "Coupling between a coherent structure and fine-scale turbulence," Phys Rev E **48**(4), 2669–2689 (1993).

[32] G.I. Taylor, "Instability of jets, threads, and sheets of viscous fluid," in *Applied Mechanics: Proceedings of the Twelfth International Congress of Applied Mechanics, Stanford University, August 26–31, 1968*, (1969), pp. 382–388.

[33] J. Buckmaster, "The buckling of thin viscous jets," J Fluid Mech **61**(3), 449–463 (1973).

[34] J.O. Cruickshank, and B.R. Munson, "Viscous fluid buckling of plane and axisymmetric jets," J Fluid Mech **113**(1), 221 (1981).

[35] M.M. Rogers, and P. Moin, "The structure of the vorticity field in homogeneous turbulent flows," J Fluid Mech **176**(1), 33 (1987).

[36] T.S. Lundgren, "Strained spiral vortex model for turbulent fine structure," Phys Fluids **25**(12), 2193–2203 (1982).

[37] R.M. Everson, and K.R. Sreenivasan, "Accumulation rates of spiral-like structures in fluid flows," Proc R Soc Lond A Math Phys Sci **437**(1900), 391–401 (1992).

[38] J.S. Marshall, "Buckling of a columnar vortex," Physics of Fluids A: Fluid Dynamics **4**(12), 2620–2627 (1992).

[39] S.E. Widnall, and D.B. Bliss, "Slender-body analysis of the motion and stability of a vortex filament containing an axial flow," J Fluid Mech **50**(2), 335–353 (1971).

[40] D. Das, M. Bansal, and A. Manghnani, "Generation and characteristics of vortex rings free of piston vortex and stopping vortex effects," J Fluid Mech **811**, 138–167 (2017).

[41] D.I. Pullin, "The large-scale structure of unsteady self-similar rolled-up vortex sheets," J Fluid Mech **88**(3), 401–430 (1978).

[42] G.K. Batchelor, *An Introduction to Fluid Dynamics* (Cambridge university press, 1967).

[43] "Improved calculations of leading-edge separation from slender, thin, delta wings," Proc R Soc Lond A Math Phys Sci **306**(1484), 67–90 (1968).

[44] N. Didden, "On the formation of vortex rings: Rolling-up and production of circulation," Zeitschrift Für Angewandte Mathematik Und Physik ZAMP **30**(1), 101–116 (1979).

[45] T. Maxworthy, "Some experimental studies of vortex rings," J Fluid Mech **81**(03), 465 (1977).

[46] D.I. Pullin, "Vortex ring formation at tube and orifice openings," Phys Fluids **22**(3), 401–403 (1979).

[47] F. and N.F. and P.T. Ducros, "Wall-adapting local eddy-viscosity models for simulations in complex geometries," Numerical Methods for Fluid Dynamics VI **6**, 293–299 (1998).

[48] I.B. Celik, Z.N. Cehreli, and I. Yavuz, "Index of Resolution Quality for Large Eddy Simulations," J Fluids Eng **127**(5), 949–958 (2005).

[49] R.K. Soni, N. Arya, and A. De, "Modal decomposition of turbulent supersonic cavity," Shock Waves **29**(1), 135–151 (2019).

[50] M. V Melander, and F. Hussain, "Core dynamics on a vortex column," Fluid Dyn Res **13**(1), 1–37 (1994).

[51] J. Jeong, and F. Hussain, "On the identification of a vortex," J Fluid Mech **285**, 69–94 (1995).

[52] C. Wu, S. Farokhi, and R. Taghavi, "Spatial instability of a swirling jet-theory and experiment," AIAA Journal **30**(6), 1545–1552 (1992).

[53] S. ARENDT, D.C. FRITTS, and Ø. ANDREASSEN, "The initial value problem for Kelvin vortex waves," J Fluid Mech **344**, 181–212 (1997).

[54] J.C.R. Hunt, "Studying turbulence using direct numerical simulation: 1987 Center for Turbulence Research NASA Ames/Stanford Summer Programme," J Fluid Mech **190**, 375–392 (1988).